\documentclass[twocolumn]{jpsj2}
\usepackage{epsfig}

\title{
Lattice Model of Sweeping Interface for
Drying Process in Water-Granule Mixture
}

\author{Hiizu \textsc{Nakanishi}, 
Ryo \textsc{Yamamoto}, Yumino \textsc{Hayase},$^1$
and Namiko \textsc{Mitarai} }

\inst{
Department of Physics, Kyushu University 33, Fukuoka 812-8581 \\
$^1$Research Institute for Electronic Science, Hokkaido
University,  Sapporo 060-0812 
}

\recdate{\today}

\abst{ 
Based on the invasion percolation model, a lattice model for the
sweeping interface dynamics is constructed to describe the pattern
forming process by a sweeping interface upon drying the water-granule
mixture.  The model is shown to produce labyrinthine patterns similar to
those found in the experiment[Yamazaki and Mizuguchi,
J. Phys. Soc. Jpn. \textbf{69} (2000) 2387].  Upon changing the initial
granular density, resulting patterns undergo the percolation transition,
but estimated critical exponents are different from those of the
conventional percolation.  Loopless structure of clusters in the
patterns produced by the sweeping dynamics seems to influence the nature
of the transition.  }

\kword{
interface dynamics, sweeping interface, pattern formation, 
invasion percolation, granular system, drying process
}

\makeatletter
\def\lsim{\mathrel{\mathpalette\gl@align<}}
\def\gsim{\mathrel{\mathpalette\gl@align>}}
\def\gl@align#1#2{\lower.6ex\vbox{\baselineskip\z@skip\lineskip\z@
    \ialign{$\m@th#1\hfil##\hfil$\crcr#2\crcr\sim\crcr}}}
\makeatother

\begin {document}
\sloppy
\maketitle
\section{Introduction}

During drying process, it often happens that material dispersed in fluid
condenses and leaves patterns on a substrate\cite{D00}.
A typical one is dry stain made by coffee spilt over a table.
Yamazaki and Mizuguchi demonstrated\cite{YM00} that the water-granule
mixture that is confined in the narrow gap between two glass plates
produces much more complicated patterns than coffee stain upon drying;
Labyrinthine patterns of granules emerge as granules are swept by
one-dimensional water-air interfaces as they recede during the drying
process.

Physical processes are as follows; As the water evaporates from the gap
between the plates, the volume of the wet region tends to shrink and the
pressure in the water decreases, which gives the driving force to move
the interface between the water and the air; Although the pressure
difference should be almost uniform over the whole interface, the
interface does not move uniformly, but only the weakest part moves at a
time.  As the interface moves, it sweeps the dispersed granules to
collect them along it; Consequently, the interface becomes more
difficult to move due to the friction of granules with the glass plates.
At some parts of the interface, the granular density eventually exceeds
the threshold where granules get stuck to form a pattern after drying.

For this phenomenon, several models have been proposed.  In their
original paper on the experiment, Yamazaki and Mizuguchi analyzed the
elementary process to estimate the thickness of the granule region
assuming the granular region as elastic material\cite{YM00}.  A phase
field model with the granular density field\cite{YMWM01,IHN05} and the boundary
dynamics\cite{N05} have been constructed, based on the observation that the
sweeping phenomenon can be regarded as the small diffusion limit of the
crystal growth\cite{L80,MS63-64}.


In this paper, focusing on similarities in dynamics between the sweeping
process described above and the invasion percolation\cite{WW83}, we
present a simple lattice model of sweeping interface for pattern
formation.  We performed numerical simulation on the model and analyzed
the resulting patterns.

\section{Lattice Model of Sweeping Interface}

The model consists of two variables at each site of the triangular
lattice:
{\em the state variable} $s_i$ and {\em the granular density} $f_i$ on
the site $i$.  The state variable $s_i$ takes the values either
0 or 1 depending
upon whether the site $i$ is dry or wet:
\begin{equation}
s_i = \left\{
\begin{array}{ll}
0 \quad & \mbox{(if the site $i$ is dry)} \\
1 & \mbox{(if the site $i$ is wet)}
\end{array} 
\right.  .
\end{equation}
The granular density $f_i$ takes zero or a positive real number
representing the quantity of granules in the $i$'th cell.

Initially, all the sites in the system are wet and the granules are
distributed uniformly with some randomness, thus the initial value of
the state variables $s_i$'s are 1 for all the sites, and the granule
density $f_i$'s are given random values from the uniform distribution
with the width $\Delta f$ and the upper bound $f_M$: $f_i\in[f_M-\Delta
f, f_M]$.


{\it The interface site} is defined as a wet site adjacent to a dry
site.  We assume the sites outside of the system are dry, thus the wet
regions are surrounded by the one dimensional chains of the interface
sites, which we call {\em the interface}.

Upon drying, the water volume shrinks, then the pressure decreases in
the water region.  The pressure difference between the wet and the dry
regions pushes the interface toward the wet region, but the site that
actually moves is the least resistive site.  The resistance, or {\em the
strength of the interface sites}, against the interface motion is
determined by the granular density $f_i$ and the local configuration of
interface;
The granules causes friction with the glass plates, while
the interface sites surrounded by the dry sites are easier to
dry due to the surface tension that tends to make the interface
straight.  In order to take these effects in a simple way, we introduce the
strength $r_i$ of the interface site $i$ as a function of its granular
density $f_i$ and the number of neighboring dry sites $n_i$. We employ
the simple form
\begin{equation}
r_i = \left\{ \begin{array}{ll}
f_i  & \mbox{if } n_i<n_s \\
f_i -\gamma  & \mbox{if } n_i\ge n_s
\end{array} 
\right.
\end{equation}
with two parameters $\gamma$ and $n_s$, which represent the surface
tension effect.

{\it Sweeping} of granules takes place when the interface moves; Upon
drying a wet interface site, the granules on this dried site are swept
away to the neighboring wet sites, as long as the granular density of
the site is not high enough for the granules to get stuck.  If the granule
density exceeds a threshold value $g_{\rm th}$, the granules get stuck
between the glass plates and cannot move, thus they stick to the
sites and form a pattern after the whole system is dried.


To represent the above processes, the dynamics for the lattice model of
the sweeping interface is defined as follows.
(i) Out of the sites with $f_i<g_{\rm th}$, pick the weakest interface
site $i$, whose strength $r_i$ is smallest.  If all the interface sites
have $f_i\ge g_{\rm th}$, then pick the weakest site from them.
(ii) Dry the site $i$ by changing its state variable $s_i$ from 1 to 0.
(iii) Redistribute $f_i$ to the neighboring wet sites equally if $f_i$
is smaller than $g_{\rm th}$ and there exist some wet
neighbor sites; Otherwise do nothing.
Repeat the steps (i) -- (iii) until all the sites becomes dry.

There are five parameters that characterize the model: $\Delta f$ and
$f_M$ to characterize the initial distribution of granular density,
$\gamma$ and $n_s$ to control the effect of surface tension on the
interface motion, and $g_{\rm th}$, that is the upper limit for the
sweeping to take place.  Out of four parameters, $\Delta f$, $f_M$,
$\gamma$, and $g_{\rm th}$, we can eliminate one that sets the scale for
the granular density;
We take $g_{\rm th}=1$.

In the following simulations, we fix the parameter $\Delta f$ to be $0.5
f_M$ for simplicity, and take $n_s=3$, i.e. half of the number of
neighboring sites in the triangular lattice that we are using.  Then we
have only two parameters, $f_M$ and $\gamma$.

\section{System behavior}

\subsection{Patterns and surface tension effects}

\begin{figure}
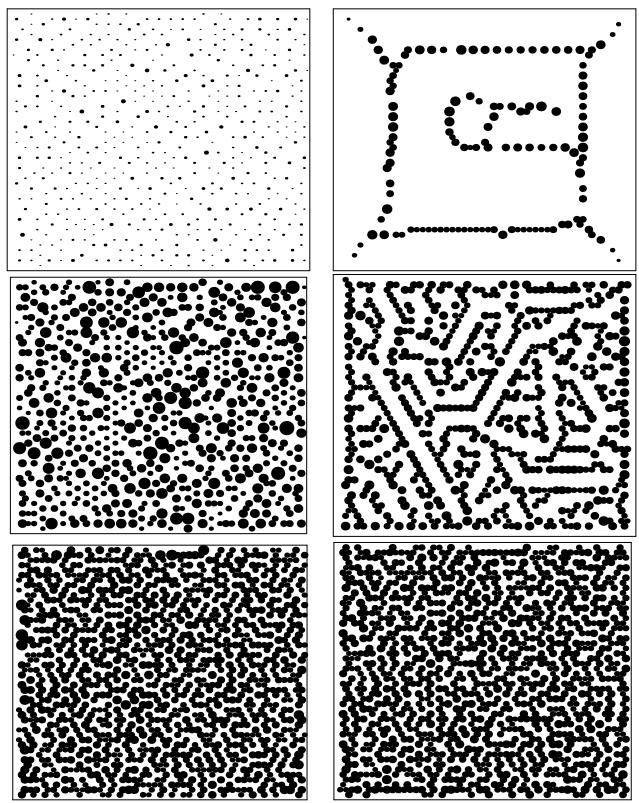

\begin{center}
\epsfig{width=3.5cm,angle=-90,file=gth1-fm01-gamma0-ns3.ps}\hskip 0.3cm
\epsfig{width=3.5cm,angle=-90,file=gth1-fm01-gamma1-ns3.ps}
\\
\epsfig{width=3.5cm,angle=-90,file=gth1-fm05-gamma0-ns3.ps}\hskip 0.3cm
\epsfig{width=3.5cm,angle=-90,file=gth1-fm05-gamma1-ns3.ps}
\\
\epsfig{width=3.5cm,angle=-90,file=gth1-fm09-gamma0-ns3.ps}\hskip 0.3cm
\epsfig{width=3.5cm,angle=-90,file=gth1-fm09-gamma1-ns3.ps}
\end{center}
\caption{ Patterns produced by the sweeping interface model with and
without the surface tension effect ($\gamma=0$ for the left column and
$\gamma/g_{th}=1$ for the right column).  The sites with the granules
are represented by the solid circles whose sizes are proportional with
the granule density.  The initial granular density is given by
$f_M=0.1$(the top row), 0.5(the middle row), and 0.9(the bottom row)
with $\Delta f=0.5f_M$ and $n_s=3$.  The linear system size is $L=64$.
} \label{Fig-1}
\end{figure}

First, we compare the patterns for $\gamma/g_{\rm th}=0$ with those for
$\gamma/g_{\rm th}=1$ in order to examine the surface tension effects in
Fig.\ref{Fig-1}, where the sites with granules are represented by the
solid circles whose sizes are proportional to the density.  The initial
grain densities are given by $f_M/g_{\rm th}=0.1$, 0.5, and 0.9, and the
system size $L\times L$ with $L=64$.  One can see the surface tension
effects especially in the lower density cases; The grains tend to align
in the systems with the surface tension $\gamma/g_{\rm th}=1$ for
$f_M=0.1$.  These correspond to the patterns obtained in the
experiment\cite{YM00}.  In the higher density cases, the surface tension
effect is small.  In the rest of the paper, we study only the case of
$\gamma/g_{\rm th}=1$ and change $f_M$.

Note that, in the case of $\gamma/g_{th}=1$, the pattern consists of
winding paths delineated by walls of granular sites; the width of
the paths are proportional to $1/f_M$, while that of the walls is always
of order of one lattice site  because we assume that granules
are swept only to the neighboring sites in the present model.

\subsection{Invading process}
\begin{figure}[h]
\begin{center}
\epsfig{width=8.5cm,angle=0,file=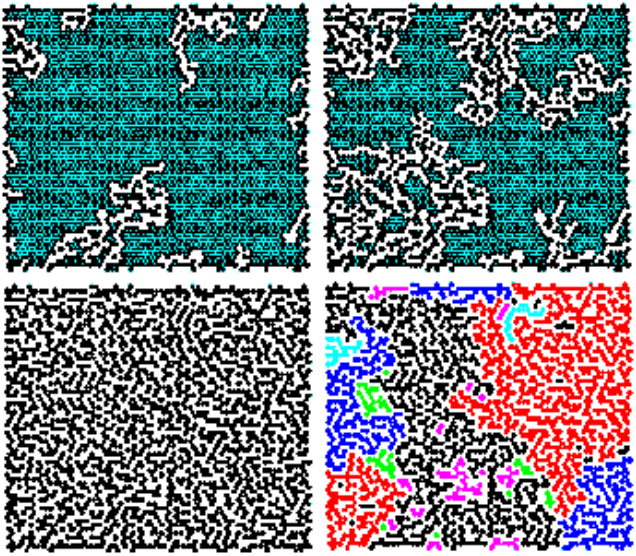}
\end{center}
\caption{
(Color online)
Time development and final patterns.
The upper two plates show the drying process. The blue
regions denote the wet regions.
The lower left plate shows the final pattern after drying with the
 granule density represented by the solid circles.
The lower right plate shows the cluster structure of the same pattern.
The parameters are
$f_M/g_{th}=0.871$, $\Delta f=0.5f_M$, $n_s=3$, $\gamma/g_{th}=1$ and $L=64$.
 } \label{Fig-2}
\end{figure}

Fig.\ref{Fig-2} shows time development of the drying process for $f_M/g_{\rm
th}=0.871$ for the system with $L=64$. While one site is dried at each
time step, the spatial development is not homogeneous but intermittent
as in the case of the invasion percolation model.  The difference in the
pattern from the invasion percolation model is that the invaded (or
dried) region develops a winding path structure, and the basic mode of
development is extending narrow paths.  This is due to the sweeping
process, by which the granules on a dried site are swept away to the
adjacent wet sites; This makes walls on both sides of the path,
consequently, it is easier to advance ahead making a narrow path than to
widen the swept region beyond a certain width.  This path extending
process leads to a labyrinthine pattern of granules when the drying
process is completed.

Both of the two plates in the lower row of Fig.\ref{Fig-2} represent the same
pattern obtained by this process; The lower left plate shows the grain
density distribution by solid circles whose size is proportional to the
density.  The pattern consists of strings of the sites with grains.  If
we define a {\it cluster} as a set of connected sites with non-zero
grain density, the whole pattern can be decomposed into clusters; In the
lower right plate of Fig.\ref{Fig-2}, its cluster structure is shown by coloring.
The clusters are of highly branched structure, but there are virtually
no loops of granular sites; The shape of clusters is compact and their
fractal dimension appears to be very close to two.  One can see that one
of the clusters is percolating from the top to the bottom; In this case,
we call the system {\it percolated}.

\section{Percolation transition of the sweeping interface model}

In the following, we will examine the percolation transition in the
pattern produced by the sweeping interface model.

\subsection{Occupation ratio}

\begin{figure}
\begin{center}
\epsfig{width=8cm,file=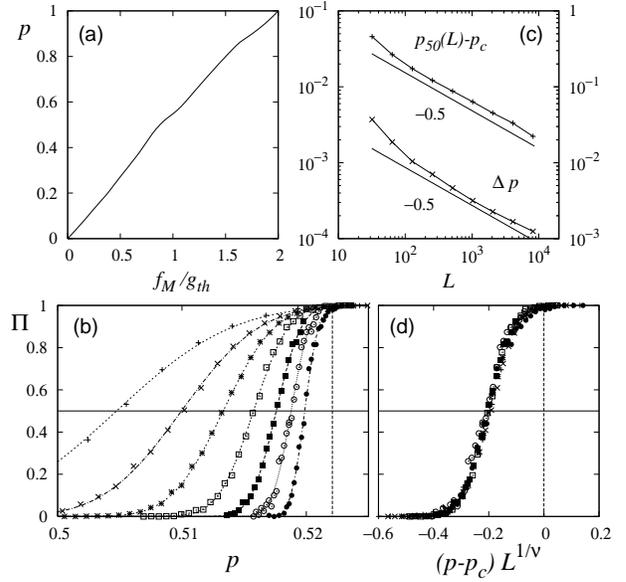}
\end{center}
\caption{
(a) Proportion of occupied sites $p$ vs. $f_M$
for the system size $L=256$.
(b) The percolation probability $\Pi$ vs. the occupation proportion $p$
for various system sizes $L$.  The system sizes are $L=$124, 256, 512,
1024, 2048, 4096, and 8192 from the left curve to the right.  The lines
denote the fitting function by the error functions with $p_{50}(L)$ and
$\Delta p$ (see text).  Each plot represents average over $200\sim 1000$
realizations.
The vertical line is drawn at $p=p_c=0.5221$.
(c) $p_{50}(L)-p_c$ and $\Delta p$ vs. $L$ in the logarithmic scale.
$p_{50}(L)-p_c$ is plotted with $p_c=0.5221$ in the left scale and $\Delta
p$ in the right scale.  The solid lines represent the lines with the
slope $-0.50$.
The other parameters of the model are $\Delta f=0.5f_M$,
$\gamma/g_{th}=1$, and $n_s=3$.
 } \label{Fig-3}
\end{figure}

The proportion $p$ of the sites with grains after drying is not an input
parameter but is obtained by performing simulations with the initial
distribution given by $f_M$.  In Fig.\ref{Fig-3}(a), $p$ is plotted
against $f_M/g_{\rm th}$.  The plot is for the system with $L=256$, but
the size dependence is small when $L$ is larger.  The ratio $p$ is an
almost linearly increasing function of $f_M/g_{\rm th}$, but one can see
a small bump near $f_M/g_{\rm th}\approx 0.9$, or $p\approx 0.5$; This
roughly corresponds with the percolation transition point.  We will not
pursue this any further in this paper although we do not understand its
nature yet.

\subsection{Percolation probability}

Fig.\ref{Fig-3}(b) shows the percolation probability $\Pi$ as a function of $p$
for various lattice size $L$.  The lines show fitting curves by the
error function as
\begin{equation}
\Pi(p;L) = {\rm Erf}\Bigl((p-p_{50}(L))/\Delta p(L)\Bigr),
\label{perco-prob-1}
\end{equation}
where the error function is defined by
\begin{equation}
{\rm Erf} (x) \equiv \frac{1}{\sqrt\pi}\int_{-\infty}^x e^{-t^2} dt ;
\end{equation}
$p_{50}(L)$ is the occupied site proportion where $\Pi=0.5$, and $\Delta
p(L)$ is the width of the percolation transition for the finite system
of the size $L$.  For lower granular density, the system is not
percolated, while the percolation probability is almost one at higher
density; The curve is steeper, or $\Delta p(L)$ is smaller, for the
larger system, which suggests that there exists a sharp transition at a
finite density in the infinite system size limit as in the case of the
conventional percolation problem.

In comparison with the conventional percolation, the difference is in the
$L$ dependence of $\Pi(p;L)$; $\Pi(p;L)$ is a decreasing function of
$L$, therefore, there is no fixed point where the percolation
probability $\Pi$'s for all $L$ intersect.

As $L$ becomes large, $p_{50}(L)$ converges to a finite value $p_c$ and
$\Delta p(L)$ goes to zero as
\begin{eqnarray}
p_c - p_{50}(L) & \sim & L^{-1/\nu_1}
\\
\Delta p(L) & \sim &  L^{-1/\nu_2}
\end{eqnarray}
with the exponents $\nu_1$ and $\nu_2\approx 1/0.50$, and the threshold
$p_c\approx 0.5221$ (Fig.\ref{Fig-3}(c)).
The corresponding value of $f_M$, or $f_c$, is
$f_c/g_{\rm th}\approx 0.926$.
  The fact that $\nu_1\approx\nu_2$ suggests that
there is a unique length scale $\xi$, which diverges as
\begin{equation}
\xi \sim |p-p_c|^{-\nu} 
\end{equation} 
with the exponent $\nu\approx 2.0$.
Note that its value 2.0 is quite different from 4/3, or the
value for the 2-d percolation\cite{SA94}.

\subsection{Order parameter}
\begin{figure}
\begin{center}
\epsfig{width=8.5cm,file=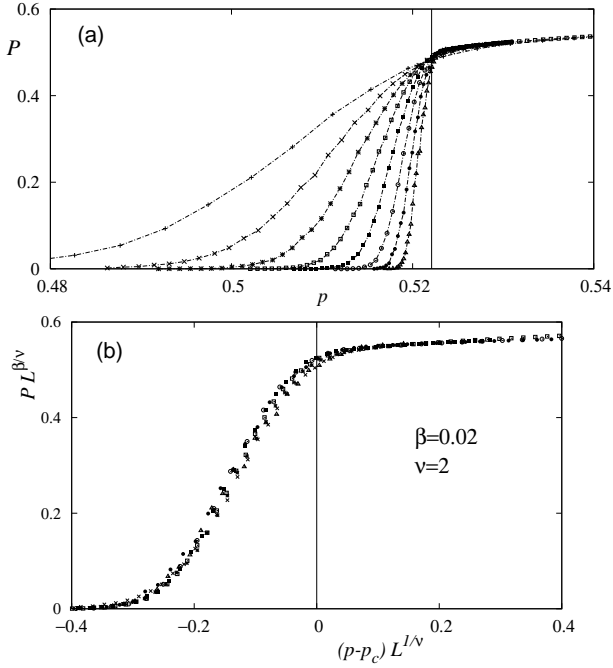}
\end{center}

\caption{
(a) The order parameter $P$ vs the occupation ratio $p$ for $L=64\sim 8192$.
The vertical line is drawn at $p=p_c=0.5221$.
(b)The scaling plot with $\beta=0.02$ and $\nu=2.0$.
 } \label{Fig-4}
\end{figure}

The percolation order parameter $P$, or the ratio of the sites that
belongs to the percolating cluster, is shown in Fig.\ref{Fig-4}(a) for $L=64\sim
8192$. The value of $P$ at $p=p_c$ is almost independent of $L$, thus the
transition looks first order, but one can observe slight tendency of
decrease in $P$ at $p=p_c$ for the systems $L> 1024$, which implies
the transition is of the second order with a small value of the exponent
$\beta$; 
\begin{equation}
P(p) \sim (p-p_c)^\beta .
\end{equation}
The scaling plots
\begin{equation}
P(p;L) = L^{-\beta/\nu} f_P\Bigl( (p-p_c) L^{1/\nu}\Bigr)
\end{equation}
with the scaling function $f_P(x)$
are shown in Fig.\ref{Fig-4}(b) with $\beta=0.02$ and $\nu=2.0$.  The plot looks
reasonably good, but one may find some scatter.

\subsection{Cluster size distribution}
\begin{figure}[h]
\begin{center}
\epsfig{width=8.5cm,file=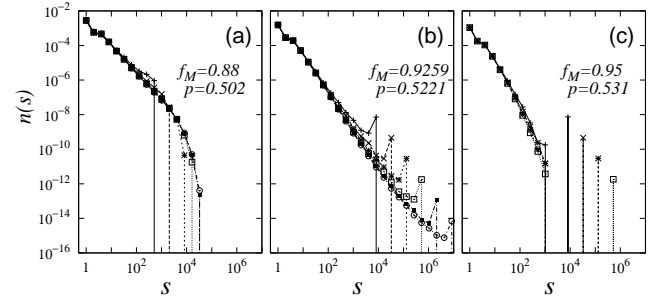}
\end{center}
\caption{ The cluster size distributions below (a), at (b), and above
(c) the percolation transition for various system sizes.  The initial
granular density is given by $f_M=0.88$(a), 0.926(b), and 0.95(c) with
the resulting $p\approx 0.502$(a), 0.522(b), and 0.531(c).  The system
sizes are $L=32\sim 1024$ (a), $128\sim 4096$ (b), and $128\sim 1024$ (c).
The other parameters are  $\Delta f=0.5f_M$,
$\gamma/g_{th}=1$ and $n_s=3$.
 } \label{Fig-5}
\end{figure}

The cluster size distribution $n(s)$ is introduced as the number of
clusters with the size $s$ divided by the number of lattice sites $L^2$
in the system, thus it is normalized as
\begin{equation}
\sum_{s=1}^{L^2} s\, n(s) = p .
\end{equation}

In Fig.\ref{Fig-5}, the cluster size distributions are shown in the logarithmic
scale for $p<p_c$ (a), $p\approx p_c$ (b), and $p>p_c$ (c).  In the case
of $p<p_c$ (a), the distribution has finite width in the large $L$ limit.
As for the case of $p>p_c$ (c), it is also of finite width but with an
isolated peak for each system size; The peak corresponds with the
percolating clusters.  The distribution at $p\approx
p_c$ in Fig.\ref{Fig-5}(b) is intriguing; They show the power law distribution
\begin{equation}
  n(s) \sim s^{-\tau}
\label{cluster-dist}
\end{equation}
with the exponent $\tau=2.25$, but there
are peaks at the large cluster end of the distribution;  The position of
the peak scales with almost $L^2$ which suggests they corresponds to the
percolating cluster.  This is consistent with the result of percolation
probability in Fig.\ref{Fig-3}(b), which shows the system with any finite size is
percolating at $p=p_c$.

\begin{figure}[b]
\begin{center}
\epsfig{width=8.5cm,angle=0,file=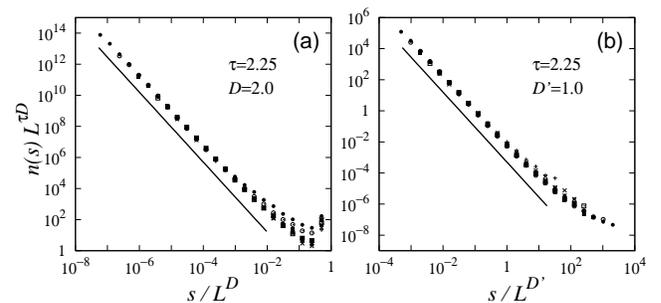}\hskip 0.2cm
\end{center}
\caption{ The scaling plot of the cluster size distribution at the
percolation transition $p=0.5221$ ($f_M=0.9259$) for $L=128\sim 8192$.
(a) The data for $s\ge 4$ are plotted with $D=2$ and $\tau=2.25$.  (b)
The data that correspond to the percolating clusters are removed in the
plot with $D'=1$ and $\tau=2.25$, using the same scaling form
(\ref{ns-scale}).  The solid lines indicate the line with the slope
$-2.25$.  } \label{Fig-6}
\end{figure}

Fig.\ref{Fig-6} shows the scaling plot
\begin{equation}
  n(s;L) = L^{-D\tau} f_n(s/L^D)
\label{ns-scale}
\end{equation}
at $p\approx p_c$ for $D=2$ (a) and 1 (b) with $\tau=2.25$ for the both.

From Fig.\ref{Fig-6}(a) with $D=2$, one can see the peak position scales
with the system area.  This is consistent with the observation that the
percolating clusters look compact and two-dimensional in
Fig.\ref{Fig-2}.  If one looks carefully, however, one may find
systematic deviation near the peak.  This can be interpreted as the
distribution deviates from the power law (\ref{cluster-dist}) at
$s>s_c(L)$, and $s_c(L)$ does not scale as $L^2$ but $L^{D'}$ with
another exponent $D'$.

To see this, we plot the same data except for the two data points that
correspond to the large percolating clusters for each $L$; We tried
various values of $D'$, but $D'=1$ seems to give least scatter around
the master curve(Fig.\ref{Fig-6}(b)).  The fact that $D$ and $D'$ are quite
different simply means that the
boundary effect has strong influence on the cluster distribution, but
the exponent $D'$ should not be associated with the fractal dimension of
the geometrical structure of clusters.

\section{Discussions and summary}



\begin{table}
\begin{tabular} {lcccc}
\hline
& $\nu$ & $\beta$ & $\tau$ & $D$
\\ \hline\hline
sweeping interface model & 2.0 & 0.02 & 2.25 & 2.0
\\ \hline
2-d percolation & 3/4 & 5/36 & 187/91 & 91/48
\\ \hline
\end{tabular}
\caption{
The critical exponents at the percolation threshold for the sweeping
interface model and the conventional 2-d percolation\cite{SA94}.
}
\label{tb-I}
\end{table}

The exponents obtained above are listed with those for the
two-dimensional conventional percolation transition(Table \ref{tb-I});
They are clearly different, suggesting that the nature of the transition
is different from that of the conventional percolation.  
This may not be very surprising because the labyrinthine patterns
analyzed here are results of the dynamical process of sweeping, which
could induce long range correlation, while the conventional percolation
transition is for the pattern by random deposition.

Most remarkable difference, however, is the behavior of the percolation
probability $\Pi(p;L)$ in Fig.\ref{Fig-3}(b).  For the sweeping
interface model, the $\Pi(p;L)$ is a decreasing function of $L$ for any
$p$, thus $\Pi(p;L)$'s for different $L$ do not intersect, while those
for the percolation intersect at $p=p_c$.  This feature, however, may be
explained by the fact that the clusters produced by the sweeping
dynamics have branching structure with virtually no loops; Even in a
very large cluster, the path that connects any given two sites in the
cluster is essentially unique\cite{com1}, thus, for a fixed $p$, the
percolating cluster is more difficult to appear in a larger system
because breakage of a path at any site during drying process may break
the cluster apart.  One consequence of this feature is that the finite
system is almost always percolated at the percolation threshold $p=p_c$
for the infinite system.

\begin{figure}[h]
\begin{center}
\epsfig{width=8.5cm,angle=0,file=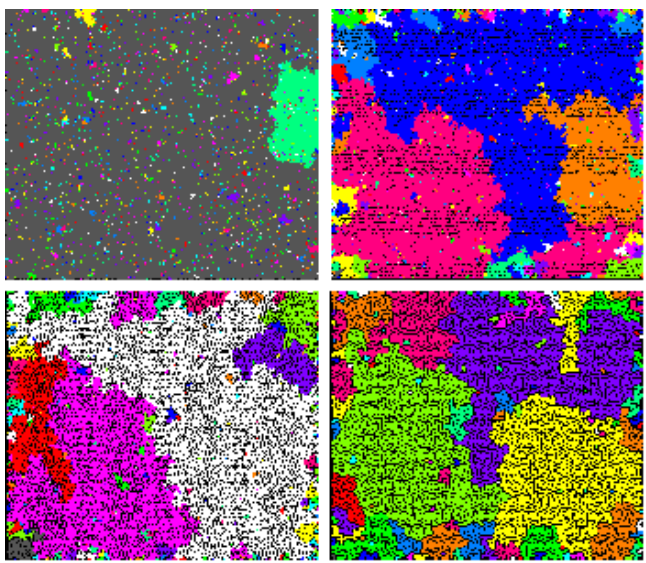}
\end{center}
\caption{ (Color on line) Cluster structure at the percolation threshold
$p=0.522$.  The cluster structure of the whole system of the size
$L=1024$ (upper left), and its subsets at the central part of the size
512 (upper right), 256 (lower left), and 128 (lower right) are shown.  }
\label{Fig-7}
\end{figure}

Another aspect of the loopless structure of cluster is that the cluster
structure of the system changes drastically when the system is trimmed.
Fig.\ref{Fig-7} shows the cluster structure of the system with $L=1024$ and its
subsystems for $p\approx p_c$.  The upper left plate is the original
system of $L=1024$.  The largest percolating cluster colored by grey is
dominating the system.
%
%
The upper right, lower left,
and lower right plates show the central parts of the original system of
the size $L=512$, 256, and 128, respectively.  The cluster structure are
analyzed only within the subsystems, {\it i.e.} the clusters that are
connected only outside of the subsystems are disconnected in the
subsystem.  One can see there are no percolating clusters in the
smallest subsystem because the original percolating cluster is
fragmented into small pieces when we look at only a smaller part of it.
This should be a general feature for the labyrinthine structure; There
is virtually no loop in clusters, therefore, there is usually only one
connecting path between any two sites in a cluster.  Consequently,
clusters are easily fragmented into pieces when they are trimmed to fit
into a smaller part of the system.

This may pose some problems in the analysis of experimental data; In the
experiment, it should be difficult to analyse whole systems because the
regions near the boundary are usually disturbed.  In the finite size
scaling analysis, the samples with smaller sizes are generated by
trimming larger samples\cite{YM00}, but the present analysis suggests
that such a procedure changes the cluster structure and results may well
depend upon the size of the original system.


In summary, being stimulated by the labyrinthine pattern produced in the drying
process, we have constructed the lattice model of the sweeping interface
based on the invasion percolation, and demonstrated the model produces
similar patterns.  The cluster analysis of the resulting patterns
shows that there is a percolation transition upon changing the density
in the initial state, but the loopless structure of the labyrinthine
patterns obtained by the sweeping dynamics make the nature of the
transition different from the conventional percolation.


\acknowledgment The authors thank Dr. Yamazaki for showing unpublished
experimental data.  This work is partially supported by a Grant-in-Aid
for scientific research (C) 16540344 from JSPS.



\begin{thebibliography}{99}
\bibitem{D00}
R.D. Deegan, Phys. Rev. E \textbf{61} (2000) 475.
\bibitem{YM00}
Y. Yamazaki and T. Mizuguchi, J. Phys. Soc. Jpn. \textbf{69} (2000) 2387.
\bibitem{YMWM01}
Y. Yamazaki, M. Mimura, T. Watanabe, and T. Mizuguchi, unpublished.
\bibitem{IHN05}
T. Iwashita, Y. Hayase, and H. Nakanishi,
J. Phys. Soc. Jpn. \textbf{74} (2005) 1657.
\bibitem{N05}
H. Nakanishi, arXiv: cond-mat/0508622,
to be published in Phys. Rev. E (2006).
\bibitem{L80}
For review, see for example,
J.S. Langer, Rev. Mod. Phys. \textbf{52} (1980) 1.
\bibitem{MS63-64}
W.W. Mullins and R.F. Sekerka, J. Appl. Phys. \textbf{34} (1963) 323;
ibid.  \textbf{35} (1964) 444.
\bibitem{WW83}
D. Wilkinson and J.F. Willemsen,
J. Phys. A {\bf 16} (1983) 3365.
\bibitem{SA94}
D. Stauffer and A. Aharony,
``Introduction to Percolation Theory'' 2nd edition, Taylor and Francis, 
1994.
\bibitem{com1}
Large loops may appear for the case 
where the average initial granular density is greater than $g_{\rm th}$,
	but near the percolation transition threshold, there exist only
	small loops
if any, because, in order that a large loop is to appear, a large area
	should be surrounded by the interface sites whose granular
	densities are all greater then or equal to $g_{\rm th}$.
\end{thebibliography}
\end {document}